\documentclass[%
singlecolumn,
preprint,
superscriptaddress,
amsmath,amssymb,
aps,
]{revtex4-1}

\usepackage{notes2bib}
\usepackage{graphicx}
\usepackage{dcolumn}
\usepackage{bm}
\usepackage{color}
\usepackage{MnSymbol}
\usepackage{romannum}

\definecolor{mygreen}{rgb}{0,0.85,0.45}
\definecolor{myred}{rgb}{1.0,0.44,0.41}
\definecolor{mypurple}{rgb}{0.57,0.49,0.84}
\definecolor{myblue}{rgb}{0,0.53,0.77}
\definecolor{mygreen}{rgb}{0,0.75,0.75}

\begin{document}


\title{Bags mediated film atomization in a cough machine}

\author{Pallav Kant}
\affiliation{Physics of Fluids Group, Max Planck Center Twente for Complex Fluid Dynamics, University of Twente, 7500 AE Enschede, The Netherlands}
\email{p.kant@utwente.nl}

\author{Cesar Pairetti}
\affiliation{Centro de Investigaci\'on en Mec\'anica Computacional (CONICET - UNL), Santa Fe, Argentina\\ and Facultad de Ciencias Exactas, Ingenier\'ia y Agrimensura (UNR), Rosario, Argentina}
\affiliation{Sorbonne Universit\'e and CNRS, Institut Jean Le Rond d'Alembert, UMR 7190, Paris, France}

\author{Youssef Saade}
\affiliation{Physics of Fluids Group, Max Planck Center Twente for Complex Fluid Dynamics, University of Twente, 7500 AE Enschede, The Netherlands}

\author{St\'ephane Popinet}
\affiliation{Sorbonne Universit\'e and CNRS, Institut Jean Le Rond d'Alembert, UMR 7190, Paris, France}

\author{St\'ephane Zaleski}
\affiliation{Sorbonne Universit\'e and CNRS, Institut Jean Le Rond d'Alembert, UMR 7190, Paris, France}
\affiliation{Institut Universitaire de France, Paris, France}
\email{stephane.zaleski@sorbonne-universite.fr}

\author{Detlef Lohse}
\affiliation{Physics of Fluids Group, Max Planck Center Twente for Complex Fluid Dynamics, University of Twente, 7500 AE Enschede, The Netherlands}
\affiliation{Max Planck Institute for Dynamics and Self-Organization, Am Fa\ss berg 17, 37077 G{\"o}ttingen, Germany}
\email{d.lohse@utwente.nl}



\begin{abstract}

We combine experiments and numerical computations to examine underlying fluid mechanical processes associated with bioaerosol generation during violent respiratory manoeuvres, such as coughing or sneezing.
Analogous experiments performed in a cough machine -- consisting of a strong shearing airflow over a thin liquid film, allow us to illustrate the changes in film topology as it disintegrates into small droplets.
We identify that aerosol generation during the shearing of the liquid film is mediated by the formation of inflated bag-like structures.
The breakup of these bags is triggered by the appearance of retracting holes that puncture the bag surface.
Consequently, the cascade from inflated bags to droplets is primarily controlled by the dynamics and stability of liquid rims bounding these retracting holes.
We also reveal the key role of fluid viscosity in the overall fragmentation process. 
It is shown that more viscous films when sheared produce smaller droplets.

\end{abstract}

\maketitle

\section{Introduction}
Respiratory droplets (bioaerosols) expelled from oral and nasal cavities during different breathing manoeuvres, such as speaking, singing, coughing and sneezing, are primary vectors of respiratory infections \cite{wells1936air, mittal2020flow, xie2009exhaled, jones2015aerosol, fiegel2006airborne, bourouiba2016sneeze, bourouiba2020turbulent, pohlker2021respiratory}.
The size of these droplets, made of muco-salivary fluid lining respiratory tracts, typically ranges from hundreds of nanometers to a few millimetres \cite{duguid1946size, loudon1967relation, chao2009characterization, bagheri2021exhaled}.
Crucially, the generation of bioaerosols is closely related to interfacial phenomena emerging at the free surface of the airway-lining-fluid \cite{evrensel1993viscous, kataoka1983generation, moriarty1999flow, pairetti2021shear}.
For example, the shearing action of the airflow within the lungs during coughing or sneezing destabilizes the thin layer of mucus accumulated along the airway passages.
Consequently, the ensuing complex fluid mechanical processes, bearing several interfacial instabilities \cite{mittal2020flow}, lead to the disintegration of the mucus layer into droplets with broad size statistics.
In addition, a large number of sub-micron-sized droplets are also produced due to the capillary breakup of menisci pending the re-opening of collapsed terminal bronchioles during normal breathing \cite{johnson2009mechanism, almstrand2010effect, malashenko2009propagation}.
Notably, the rheological properties of the muco-salivary fluid play a crucial role in these atomization processes, thus, dictating size statistics of droplets exhaled \cite{hamed2020surface, keshavarz2016ligament}.
Upon exhalation, the fate of these respiratory droplets largely depends on their size.
While, large droplets of size greater than 100 $\mu$m behave ballistically due to their own weight, falling out of the exhaled turbulent ``puff'', smaller droplets often referred to as \emph{nuclei} remain airborne for long periods.
The lifetime of these airborne droplets then depends on a number of intrinsic factors such as intial size and composition, as well as extrinsic factors like ambient humidity, temperature and ventilation \cite{morawska2006droplet, holmgren2011relation, villermaux2017fine, chong2021extended}.
Therefore, devising an effective infection control strategy, first and foremost, requires an intimate knowledge of the initial size distribution, velocity and ejection angles of bioaerosols generated during the different respiratory manoeuvres. 

Here, in the context of droplets produced during coughing or sneezing, we study the fragmentation of thin films subject to interfacial stresses induced by a strong airflow.
We employ the experimental configurations popularized by Clark \emph{et al.} \cite{clarke1970resistance} and King \emph{et al.} \cite{king1985clearance} to study the flow resistance in airway passages due to the accumulation of mucus along the walls.
Note that the scope of these earlier investigations was limited to non-atomization conditions in which modal stability analysis could explain observed interfacial disturbances.
More recently, Edwards \emph{et al.} \cite{edwards2004inhaling} employed a similar apparatus, along with \emph{in vivo} measurements, to establish a direct link between the interfacial tension of the airway-lining fluid and the flux of exhaled bioaerosols.
In contrast, here we use high-speed imaging and Phase-Doppler-Anemometry (PDA) to reveal the mechanics of fragmentation of the thin films and to characterize the resultant droplet size distributions, respectively.
Additionally, we present a state-of-the-art numerical framework that can be used to study a similar class of fluid fragmentation problems.

\section{Methods}

\subsubsection{Experimental Setup}

\begin{figure}[htb]
\centering
\includegraphics[clip, trim=0cm 0cm 0cm 0cm, width=0.6\textwidth]{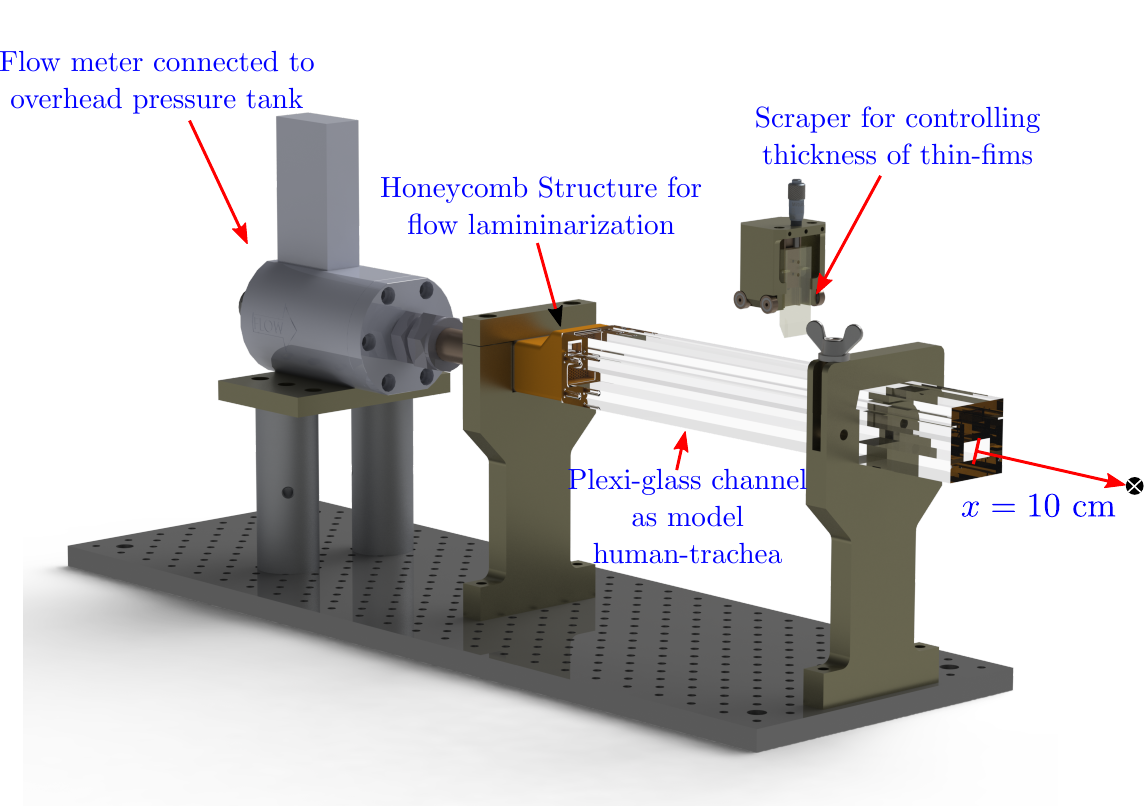}\\
\caption{Schematic diagram of the experimental setup. A transparent plexi-glass channel with rectangular cross-section (model-trachea) is connected with a pressurized tank via a valve. A thin layer of aqueous glycerol solutions covers the bottom of the channel. Upon actuation of the valve, the events leading to the fragmentation of thin-film are recorded by a high-speed camera in top view. Droplet statistics are measured at a distance of 10 cm away, along the center-line, from the exit of the channel.}
\label{fig:setup}
\end{figure}

We perform analogous experiments and numerical computations of what we call a \emph{cough-machine} to illustrate the topological changes of a thin film as it disintegrates into minute droplets.
A schematic diagram of the experimental setup is shown in Fig.\,1.
The experimental configuration (\emph{cough-machine}) consists of a $30\,\mathrm{cm}$ long, horizontal model-trachea made of transparent plexi-glass, with a $2\,\mathrm{cm}$ wide and  $1\,\mathrm{cm}$ high rectangular cross-section.
A thin, highly wetting glass coverslips of thickness 100 $\mu$m and length 12 cm cover the bottom of this rectangular channel at the upstream.
The wetting nature of the coverslip allows us to create a uniform layer of an aqueous glycerol solution on top of it, imitating a coating of a muco-salivary fluid in the trachea, away from the oral cavity.
The wettability of these removable coverslips was controlled by treatment with plasma after every few experiments.
The desired film thickness $\mathcal{H}_\mathrm{f}$ is achieved by first layering the bottom of the cough machine with a fixed volume of glycerol solution, and then spreading it uniformly with a sharp edge scraper that runs over a rail.
Experiments are carried out with different aqueous glycerol solutions of viscosity ($\mu_\mathrm{l}$) ranging between 1 mPa\,s  and 132 mPa\,s.
This range matches with the typical viscosity measurements reported for human mucus \cite{lai2009micro}.
Note that the typical cross-sectional shape of the tracheal lumen in an adult is ovoid \cite{furlow2018surgical}, however, here we chose a rectangular cross-section to achieve a uniform film of the model muco-salivarly fluid, which in the case of a round tube would tend to collect at the bottom. 
The upstream of the channel connects to a pressurized tank via a valve that is actuated to create an airflow mimicking a typical cough/sneeze, and the other end opens to atmosphere.
A honeycomb structure is placed at the entrance of the channel to homogenize the inlet turbulence levels.
During an experiment the valve opens for 400 ms.
Within this duration the flow rate quickly ramps-up to its maximum and then drops down like a real cough \cite{king1985clearance}.
We must emphasize that the atomization of a thin film, as discussed in the following, takes place at much shorter time-scales $\sim 1 \mathrm{ms}$ compared to the duration of operation of the valve (400 ms).
Thus it is valid to consider approximately constant and uniform flow conditions during an atomization event.
Different free-stream velocities ($U \sim $ 10 - 30 m/s) within the channel are achieved by varying the overhead pressure in the tank.
Therefore, the corresponding shearing strength of the airflow ranges in between $\Gamma \sim \mu_\mathrm{l}\,U/\mathcal{H}_\mathrm{f} \sim 10 - 3900\,\mathrm{Pa}$.
Note that the mean free-stream velocity is determined from the flow-rate measured at the inlet of the channel via a flow meter (Bronkhosrt, EL-Flow Select).
The atomization of thin-film inside the channel is recorded in top-view via a high-speed camera (SA12-NOVA, Photron ) at a maximum of 40,000 frames per second, and droplet size statistics are measured, using a Phase-Doppler-Anemometer (Dual PDA, Dantec Dynamics) at a distance of 10 cm from the exit of the channel as shown in Fig\,1.
For details on the underlying principles of PDA technique, we refer the reader to Ref.\cite{qiu1992reliable}.


\subsubsection{Numerical scheme and Simulation setup}
In the present investigations, both the muco-salivary fluid and the surrounding gas are treated as incompressible Newtonian fluids. 
In the context of uniform temperature conditions achieved during experiments, and high relative humidity within the expelled multi-phasic puff from the channel, heat transfer and evaporative effects are ignored \cite{chong2021extended, villermaux2017fine}. 
Accordingly, the flow in the gas medium and the liquid film is described by the Navier-Stokes equations as:
\begin{equation}
\begin{gathered}
\nabla \cdot \vec{u} = 0,\\
\frac{\partial \rho \vec{u}}{\partial t} + \nabla \cdot \left(\rho\,\vec{u}\,\vec{u}\right) = -\nabla p + \nabla \cdot \left(2\,\mu\,\mathbf{D}\right) + \sigma\,\kappa\,\vec{n}_s\,\delta_s,
\end{gathered}
\label{eq:NS}
\end{equation}
where $\vec{u}\left(\vec{x},t\right)$ is the velocity field and $p\left(\vec{x},t\right)$ is the pressure field. 
Tensor $\mathbf{D}$ is equal to $\frac{1}{2}\left[ \nabla \vec{u} + \left( \nabla \vec{u} \right)^{\mathrm{T}} \right]$. $\rho$ and $\mu$ are the flow density and viscosity respectively. 
The last term on the right-hand side in Eq.(\ref{eq:NS}) represents the surface tension force, where the coefficient $\sigma$ is the interfacial tension between the aqueous glycerol solution and air. 
The force only acts at the free surface, hence the Dirac function $\delta_s$, and also depends on the interface shape, particularly on its curvature $\kappa$ and normal $\vec{n}_s$

Simulations are performed for flow in a rectangular channel with bottom wall covered by a thin liquid film of thickness $\mathcal{H}_\mathrm{f}$ as described in the previous section.
No-slip boundary condition is imposed on all dry boundaries of the channel.
At the inlet cross-section, we apply uniform velocity, and outflow boundary conditions of Dirichlet on pressure and Neumann on velocity are imposed on the outlet cross-section.
We solve the above equations by a simple finite volume discretization in the one-fluid numerical approach for two-phase flows, using a Piecewise Linear Interface Capturing (PLIC), Volume-Of-Fluid (VOF) method. 
The VOF function at a grid cell is defined as $f = \left( 1/V \right)\,\int_V c\,\mathrm{d} V$, representing the volume fraction of the liquid phase in a grid cell.
Here, $c$ indicates the presence of the liquid $\left(c \left(\vec{x},t\right) = 1\right)$ or $\left(c \left(\vec{x},t\right) = 0\right)$ gas phase. 
The fluid properties, density ($\rho$) and viscosity ($\mu$), at the cell are defined by arithmetic averages:
\begin{equation}
\rho = f\rho_\mathrm{l} + \left( 1 -f\right) \rho_\mathrm{g} \quad \mu = f\mu_\mathrm{l}+ \left( 1 -f\right) \mu_\mathrm{g},
\end{equation}
subscripts `$\mathrm{l}$' and `$\mathrm{g}$' correspond to the liquid and gas properties, respectively.
The flow equations are discretised using the methodology described in Ref.\,\cite{pairetti2021shear}.
Notably, the advective fluxes in the discretized system are integrated explicitly based on a momentum-conserving Bell-Collela-Glaz scheme \cite{bell1989second} and the viscous term is computed in a semi-implicit manner. 
The surface tension force term is based on a discrete-balanced formulation to reduce spurious currents\cite{popinet2018numerical}. 
Importantly, we compute the interface curvature using second-order height functions defined by PLIC plane position based on the analytical formulation from Ref.\,\cite{scardovelli1999direct}
The described flow solver is implemented in the open-source software program Basilisk \cite{basilisk}.
In the simulations reported here, we apply Adaptive Mesh Refinement (AMR), in an octree mesh implementation using a wavelet-based criterion \cite{van2018towards}.
The finest simulation reported employs a maximum refinement level of thirteen (L13), resulting in a grid size of $18.3\, \mu$m.
For more details on the numerical scheme we refer the reader to Ref.\,\cite{pairetti2021shear}.

\section{Results and Discussion}
\begin{figure*}[htb]
\centering
\includegraphics[clip, trim=0cm 0cm 0cm 0cm, width=1.0\textwidth]{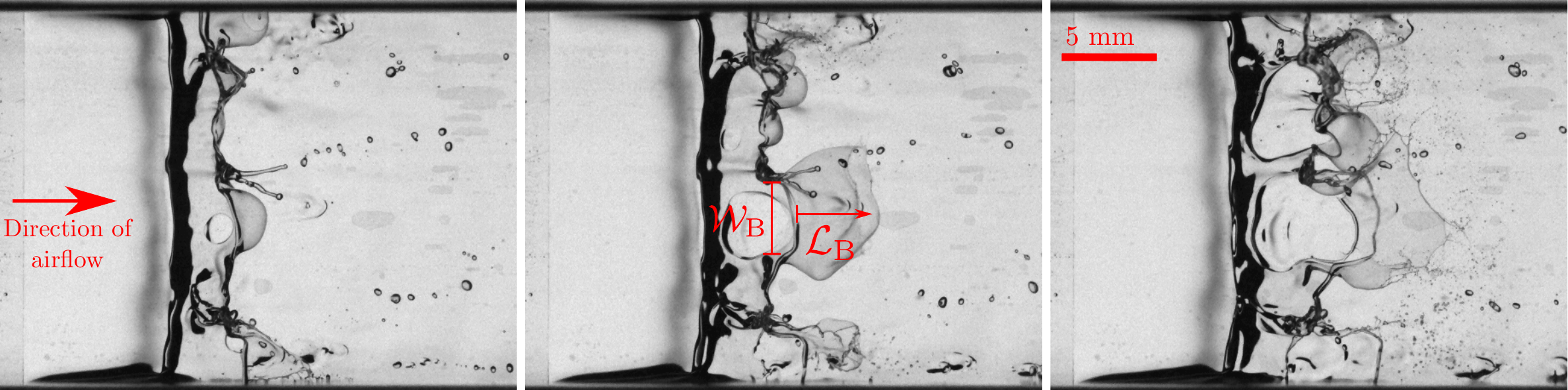}\\
\caption{Sequence of experimental snapshots, at 0.25 ms intervals, highlight the key processes during the fragmentation of a thin-film of water-glycerol solution (film thickness $\mathcal{H}_\mathrm{f} = 1$ mm and kinematic viscosity $\nu_\mathrm{l} = 5$ cst) when sheared by an impulsively started airflow in a cough-machine. The mean velocity of the airflow is 15 m/s. The sequence shows the topological changes experienced by the film as multiple bags form along its width. The disintegration of these inflated structures generates droplets with broad size statistics. The first snapshot is captured 40 ms after the initiation of airflow in the channel.}
\label{fig:fig2}
\end{figure*}

The sequence of snapshots in Fig.\,1a highlights key events leading to the atomization of a water-glycerol film of thickness $\mathcal{H}_\mathrm{f} = 1$mm when sheared by an impulsively started airflow, mimicking a cough or a sneeze.
Since the corresponding Reynolds number $Re = \rho_\mathrm{l}\, \mathcal{H}_\mathrm{f}\,U/\mu_\mathrm{l}$ and Weber number $We = \rho_\mathrm{g}\,U^2\,\mathcal{H}_\mathrm{f}/\sigma$ are sufficiently large, owing to a shear-induced Kelvin-Helmholtz instability, a localized wave-like interfacial disturbance develops close to the inlet immediately after the airflow initiates \cite{scardovelli1999direct, hoepffner2011self}; here, $\rho_\mathrm{l}$, $\mu_\mathrm{l}$ and $\sigma$ are the density, the dynamic viscosity, and the surface tension of the liquid, $\rho_\mathrm{g}$ and $U$ are the density and the mean flow velocity of the flowing gas.
While this interfacial wave-like disturbance moves downstream at a constant velocity that scales as $u_\mathrm{wave} \sim U \sqrt{\rho_\mathrm{g}}/(\sqrt{\rho_\mathrm{g}} + \sqrt{\rho_\mathrm{l}}))$ \cite{dimotakis1986two, hoepffner2011self}, the shear stress imposed by the airflow on the thin film modulates its morphology, stretching it into a thin liquid sheet with a thicker (cylindrical) rim at its free edge.
At this stage, the liquid sheet disintegrates into small droplets with a wide size distribution primarily through the formation of multiple inflated structures (liquid bags) along the width of the film.
Notably these hollow structures remain rooted at the crest of the downstream traveling interfacial wave, and are bordered by a thick liquid rim at the top.
Since large accelerations are imparted from the gas flow to the liquid sheets during the expansion of the bags, their formation and growth are attributed to a Rayleigh-Taylor type mechanism.
Correspondingly, the width of bags at the location of attachment with the interfacial wave is prescribed by the dominant wavelength associated with the Rayleigh-Taylor instability $\lambda \sim \sqrt{\sigma/\rho_\mathrm{l} a}$, where $a \sim \mathcal{H}_\mathrm{f}/\tau^2$ is the initial acceleration of the inflating bag; $\tau = \mathcal{H}_\mathrm{f}/U \sqrt{\rho_\mathrm{l}/\rho_\mathrm{g}}$ is characteristic time scale associated with bursting of a bag \cite{villermaux2009single}.
For different glycerol solutions and experimental conditions ($\mathcal{H}_\mathrm{f}$, U) investigated in experiments, the predicted instability wavelength is of the order of 2 mm, which is in agreement with observed sizes of bags in experiments.



Corresponding 2D and 3D numerical simulations (Fig.\,3) performed for the exact experimental configuration reveal an intricate interplay between aerodynamic stresses at the film interface and capillarity that drives the inflation of liquid bags.
We find that once the amplitude of the initial interfacial perturbation increases significantly, the flow over its windward side develops a stagnation point where the pressure builds up, stretching it into a liquid sheet. 
Subsequently, local pressure gradients thins the liquid sheet heterogeneously, driving the inflation of several bags across the channel width.
Upon the initial inflation, vortex shedding develops on the downwind side producing stagnation points that accelerate the local thinning of the sheets while the bags keep elongating.

\begin{figure*}[htb]
\centering
\includegraphics[clip, trim=0cm 0cm 0cm 0cm, width=0.9\textwidth]{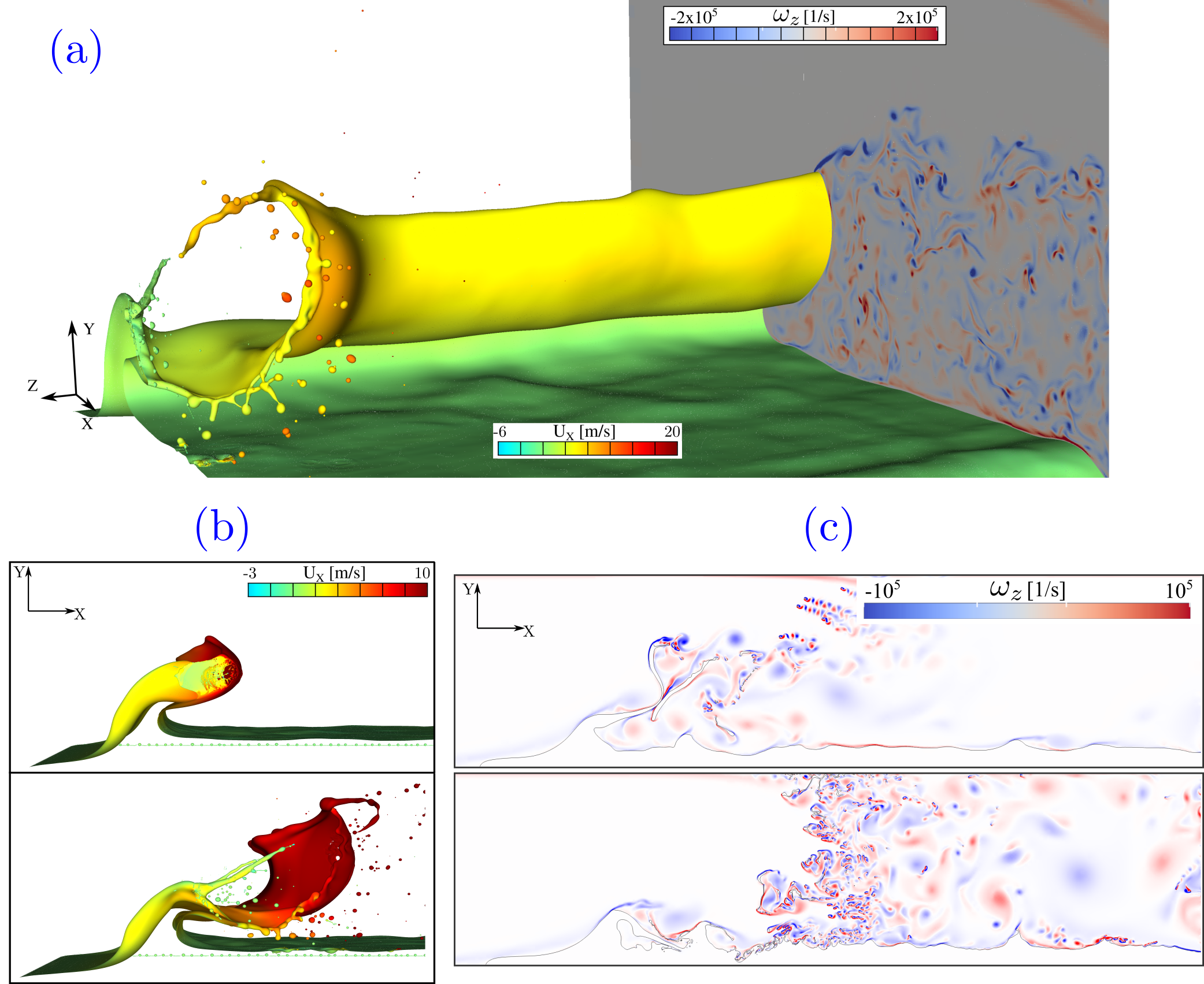}\\
\caption{(a) An isometric view of that bag morphology resulting from the shearing of a thin film of thickness $\mathcal{H}_\mathrm{f} = 1$mm within the cough machine. Side-walls of the rectangular channel (\emph{cough-machine}) are not shown for the ease of visualization. The shearing velocity used in the 3D computations is 15 m/s. A vorticity map shown on the vertical cross-section (grey panel) highlights the flow structure around the inflated liquid bag. (b) Snapshots illustrating the inflation and atomization of a liquid bag observed in 3D simulations. (c) Side-view of bag-breakup mode observed in 2D simulations. All computations are performed using \emph{Basilisk}.}
\label{fig:fig3}
\end{figure*}

Notably, both in our experiments and simulations, the bag-mediated fragmentation of a thin film occurs when the Weber number exceeds the threshold value $We_\mathrm{c} \sim 8$.
Below this threshold value we do not observe any atomization of a thin film by the shearing airflow.
Interestingly, this threshold also corresponds to the limit beyond which a droplet exposed to high-speed airflow fragments into smaller droplets via a `bag-breakup' mode \cite{villermaux2009single, reyssat2007shape, marcotte2019density, jackiw2021aerodynamic}.
This suggests that the underlying physical mechanism driving the morphological change from the initial interfacial disturbance to a hollow inflated structure, as described above, is analogous to the case of impulsively accelerated droplet inflating into a bag morphology \cite{hsiang1992near, villermaux2009single, kulkarni2014bag, guildenbecher2009secondary, reyssat2007shape, marcotte2019density, jackiw2021aerodynamic}.
In both scenarios, the aerodynamic drag on the bulk liquid first, deforms it into a thin sheet transverse to the airflow, which then inflates into a bag shape due to a stagnation-point flow on its windward face.  
Crucially, we find that the critical Weber number beyond which such bag-mediated atomization events are observed in current experiments remains unchanged for the range of fluid viscosities considered in experiments.
Understanding the influence of viscous effects on critical Weber number for bag formation on sheared thin films demands a detailed investigation addressing the non-linear stretching and thinning of the interfacial wave formed initially due to Kelvin-Helmholtz instability, which is beyond the scope of the current investigation.
However, we should point out that this weak dependence of the critical Weber number on film viscosity is well captured by the theoretical prediction  $We_\mathrm{c} = 12 \left(1+ (2/3)\mathrm{Oh}^2 \right)$ \cite{kulkarni2014bag} suggested for bag inflation from a droplet, which accounts for viscous effects through Ohnesorge number $\mathrm{Oh} = \mu_l/(\rho_l \sigma \mathcal{H}_\mathrm{f})^{1/2}$.
Note that for the range of film viscosities used in experiments the maximum Ohnesorge number is of the order of $10^{-1}$.
Accordingly, theoretical predictions of the critical Weber number show only minor variations for different film viscosities employed in experiments.

Using top-view recordings, we quantify the kinematics of the longitudinal and transversal expansion of liquid bags observed in current experiments.
The data shown in Fig.\,4a shows the longitudinal expansion of bags measured in experiments, for different film viscosities, as well as the corresponding behaviour recovered in numerical simulations.
Remarkably, similar to the bag inflating from an impulsively accelerated droplet, we find that the temporal growth of the horizontal position of the tip of a bag in our experiments follows the power-law like behaviour  $\mathcal{R}_\mathrm{Tip} \sim t^2$ \cite{villermaux2009single}; $\mathcal{R}_\mathrm{tip}$ is the position of the tip of an inflating bag.
This suggests that the the inflation of a bag is primarily dictated by inertial effects, while viscous effects play a dormant role.
Moreover, in our numerical simulations we also recover the exponential thinning of the bag thickness $h \sim e^{-2t}$ during the inflation process, also suggested for the single droplet case \cite{villermaux2009single}.
These comparisons confirm the strong similarity in physical mechanisms that dictate the bag-mediated fragmentation process in the two distinct scenarios.
However, the transverse expansion of the inflated morphologies observed in our experiments shows an interestingly different trend.
Since the bags in current experiments always remain rooted at the crest of the downstream traveling interfacial wave, unlike the bags that are unhinderedly inflating all directions from a droplet, the transverse growth of bags is significantly altered.
We find that the width of a bag at the location of attachment with the crest of the interfacial wave $\mathcal{W}_\mathrm{B}$ remains nearly constant throughout its lifetime.
As highlighted in Fig.\,4c, after an initial rapid growth at the root, that lasts less than 0.1 ms, the width of the bag does not change. 
In addition, we find that the maximum width of the bag is significantly affected by the viscosity of the thin-film.
We observe that at a fixed Weber number, wider bags are favored for more viscous films; see Fig.\,4d. 
Since the formation of the inflated morphologies is attributed to a Rayleigh-Taylor instability type mechanism, this observed trend in the width of the bag is in line with the expected increase in the dominant unstable wavelength with viscosity \cite{plesset1974viscous}.
At the same time, we must also draw attention towards the apparent plateau in measured bag-widths at higher viscosities.
We suspect this behaviour emerges from the physical constraint on the largest unstable wavelength that the stretched interfacial disturbance on the thin film can accommodate.
A detailed investigation is further warranted to address this crucial aspect of the system, which is beyond the scope of the current investigations.

\begin{figure}
\centering
\includegraphics[clip, trim=0cm 0cm 0cm 0cm, width=0.9\textwidth]{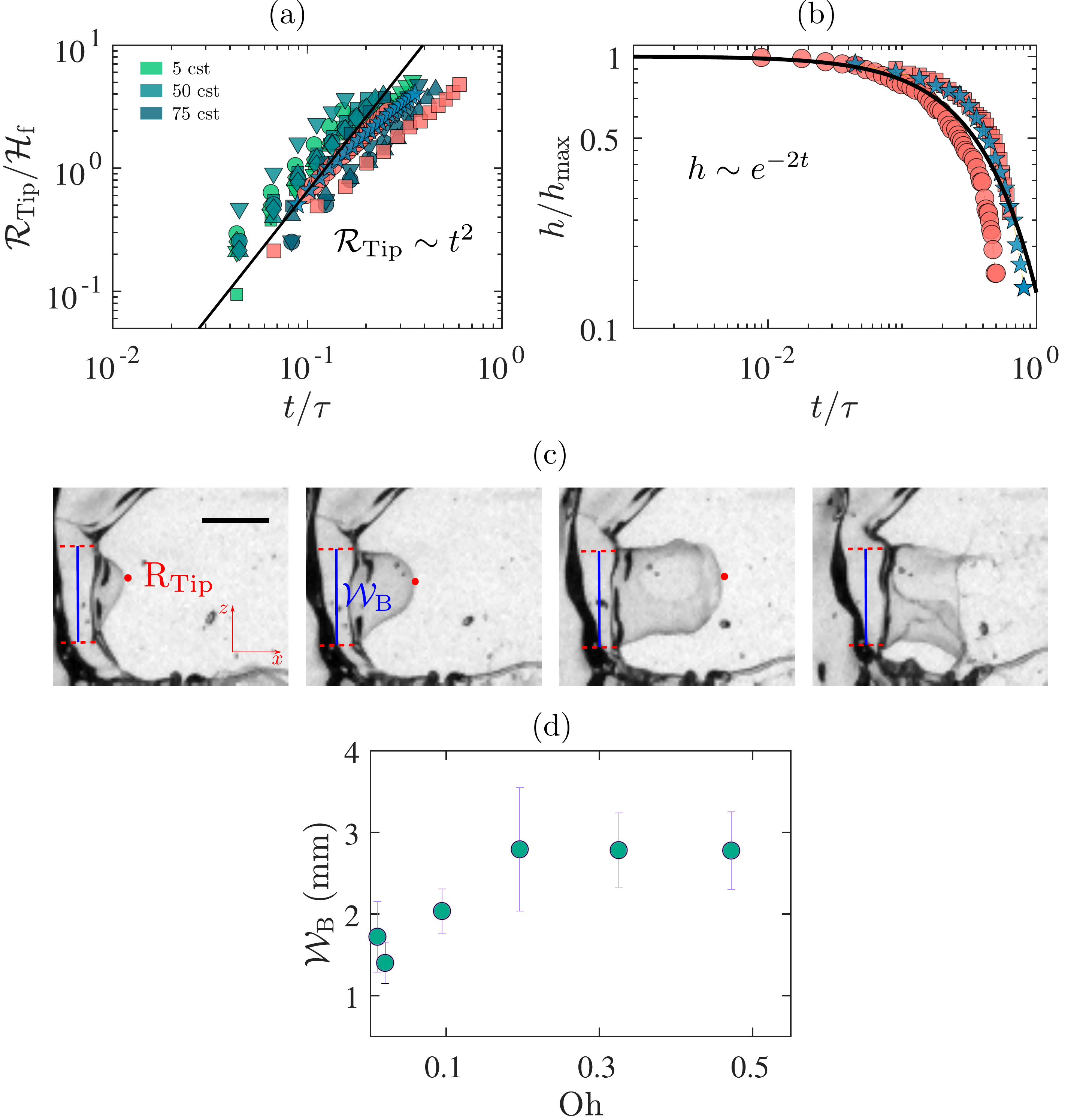}\\
\caption{(a) Kinematics of bag expansion measured for different film viscosities $\nu$ = 5 cst, 50 cst and 75 cst. Symbols are color coded according to the film viscosity. Data from 3D simulations (\textcolor{myblue}{$\filledlargestar$} = 15\,m/s) and 2D numerical simulations (\textcolor{myred}{$\bullet$} = 30\,m/s and \textcolor{myred}{$\filledmedsquare$} = 15\,m/s) are performed for film viscosity $\nu$ = 5 cst. Solid black line is theoretical prediction for bag-inflation kinematics, $\mathcal{R}_\mathrm{Tip} \sim t^2$. (b) Theoretical exponential decay in bag thickness, $h \sim e^{-2t}$ (solid black line), measured in numerical computations. Here, the bag-thickness is non-dimensionalised by its maximum value. (c) Sequence of experimental snapshots illustrating the fact that the width of the bag $\mathcal{W}_\mathrm{B}$ at the location of the attachment with the crest of the interfacial wave remains nearly constant while it inflates. The vertical lines on each snapshot corresponds to the instantaneous width of the bag. The horizontal scale-bar shown on the first snapshot corresponds to 1 mm. (d) Experimentally measured variation in bag width with film viscosity. The data corresponds to experiments performed with film thickness $\mathcal{H}_\mathrm{f} = 1\,\mathrm{mm}$ and $U = 30\,\mathrm{m/s}$. }
\label{fig:fig4}
\end{figure}

High-speed imaging of the fragmentation of a bag reveals that its fragmentation into small droplets is triggered by the appearance of `weak-spots', particularly in the thinnest parts of the bag-sheet \cite{villermaux2020fragmentation, lohse2020double}. 
As shown in Fig.\,5, these weak-spots are precursor to retracting holes that puncture the bag.
These holes are bounded by thick liquid rims and retract at a constant velocity.
The corresponding retraction thus is controlled by a Taylor-Culick process \cite{savva2009viscous,  villermaux2009single, lhuissier2012bursting}, and the retraction velocity is given by $V_\mathrm{TC} = (2\sigma/\rho_\mathrm{l}\,h)^{1/2}$; $h$ is the thickness of the punctured liquid sheet.
We use this relationship between the retraction velocity and the liquid sheet thickness to estimate the typical bag-thickness at the time of their bursting in our experiments.
Our calculations indicate that bursting events occur when the bag-thickness is close to 1 $\mu$m.
Additionally, we find that viscous effects become important in the final moments of the bag inflation, thus controlling the minimum thickness of the bag and the resultant droplet size statistics.
This crucial role of liquid viscosity on the droplet generation process is addressed in detail later in the discussion.

\begin{figure}
\centering
\includegraphics[clip, trim=0cm 0cm 0cm 0cm, width=0.42\textwidth]{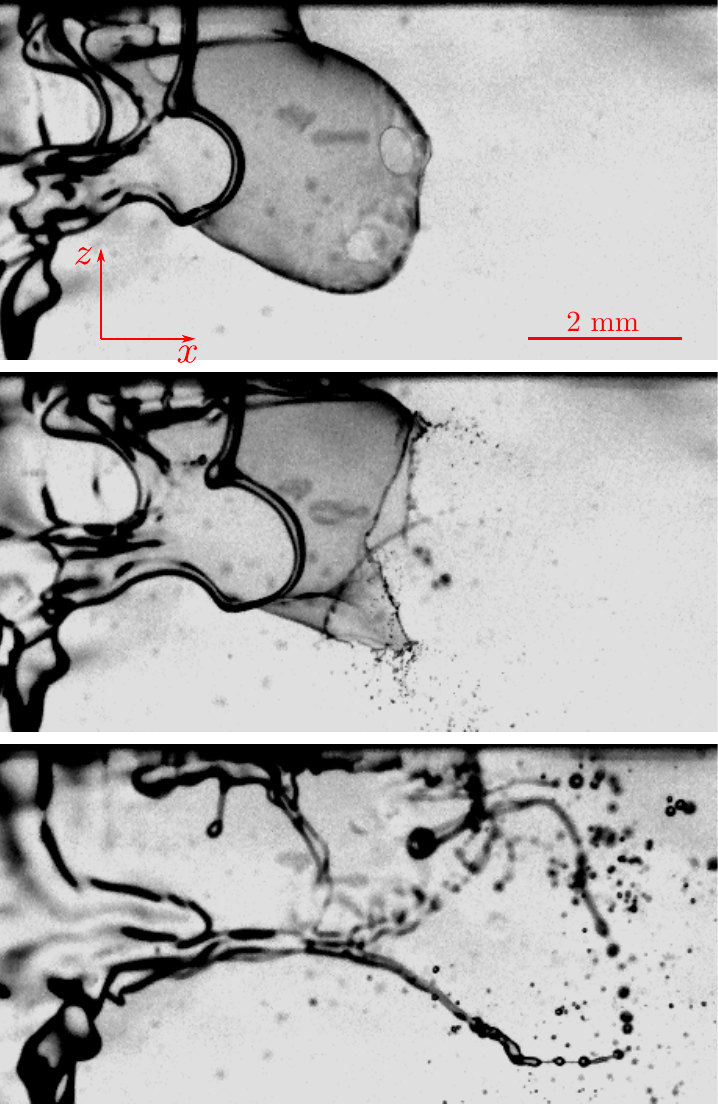}\\
\caption{The disintegration of liquid-bags is triggered by the appearance of `weak-spots'. These weak-spots are precursor of holes that puncture that bag (shown in the top frame). During the expansion of these holes, a large number of minute droplets are produced due to instability of the retracting liquid rim (middle frame). Finally, larger droplets form as the corrugated filament bordering the inflated bag breaks up (bottom frame). Experimental snapshots shown here are are 0.075 ms apart.}
\label{fig:fig5}
\end{figure}

Note that the exact physical mechanism responsible for the nucleation of such weak spots is a subject of intense debate \cite{villermaux2020fragmentation, lohse2020double, villermaux2007fragmentation}. 
Accordingly, several nucleation mechanisms relying on chemical/thermal inhomogeneities or the presence of foreign particles or micro-bubbles have been proposed. 
A few other investigations have also pointed out that extremely thin liquid sheets of thickness less than 1 nm are susceptible to thermal fluctuations or intermolecular interactions, which may also cause the nucleation of weak-spots \cite{sharma1996instability, casteletto2003stability}.
However, in the view of thicker bag sheets observed in our experiments and uniform experimental conditions, the above-mentioned mechanisms seemingly may not cause the nucleation of weak spots in our experiments.
Instead, we speculate that the nucleation of weak-spots in our experiments may occur as a result of mechanical forcing, like the application of a pressure gradient, or pressure difference across a film \cite{villermaux2020fragmentation, vledouts2016explosive, klein2020drop}. 
In such a case, the resulting acceleration may lead to film thickness modulations through a Rayleigh-Taylor mechanism, ultimately causing its puncture \cite{bremond2005bursting}.

Crucially, our experiments reveal that the cascade from liquid sheet to droplets is predominantly controlled by the dynamics and stability of liquid rims formed upon the nucleation of holes.
The retracting motion of such liquid rims is susceptible to longitudinal perturbations, resulting in the formation of elongated liquid-filament at the free edge which eventually break into small droplets \cite{agbaglah2013longitudinal, krechetnikov2010stability, roisman2010instability}.
This transition from a liquid-filament to droplet/s occurs either via end-pinching mode when a filament grows into a bulge that detaches as a single droplet, or via ligament mode when the filament grows progressively to break up into multiple droplets \cite{wang2021growth}.
In the experiments, we observe both modes of droplet generation.
Notably, this physical process of droplet generation is controlled by a delicate interplay between Rayleigh-Plateau and Rayleigh-Taylor mechanisms \cite{wang2018universal, wang2021growth}.
The size of droplets (less than 50\,$\mu$m) produced via this mechanism is therefore determined by the size of the liquid rim that continuously evolves in time.
Additionally, as shown in the last image of the sequence in Fig.\,5, droplets of larger size ($>$ 100\,$\mu$m) are also produced from the breakup of the corrugated filament at the top-edge of the liquid bag via an aggregation-coalescence process \cite{villermaux2007fragmentation}.
At this point, it is essential to point out that a significant portion of the total liquid volume that fragments into droplets is located in the inflated bag sheets $\mathcal{V}_\mathrm{sheet} \sim \mathcal{L}_\mathrm{B} \mathcal{W}_\mathrm{B} h$, and it is comparable to the total liquid volume carried by the liquid filament at edge of the bags ($\mathcal{V}_\mathrm{filament} \sim r_\mathrm{filament}^2 \mathcal{W}_\mathrm{B}$), as $\mathcal{L}_\mathrm{B} h \sim r_\mathrm{filament}^2$
Therefore, the poly-dispersity and skewness of the resultant droplet size distribution measured in our experiments is attributed to droplets originating from the bursting of inflated liquid sheets and the breakup of bordering corrugated filaments.
We must point out that a similar conclusion was also reached in a recent experimental investigation concerning bag-mediated secondary atomization of a droplet \cite{jackiw2022prediction}.

\begin{figure}
\centering
\includegraphics[clip, trim=0cm 0cm 0cm 0cm, width=0.7\textwidth]{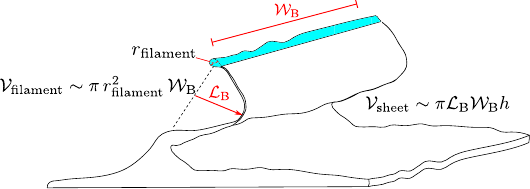}\\
\caption{A schematic comparing the volume accompanying the inflated liquid sheet and the filament bordering thee inflated morphology.}
\label{fig:fig6}
\end{figure}

\begin{figure}
\centering
\includegraphics[clip, trim=0cm 0cm 0cm 0cm, width=0.8\textwidth]{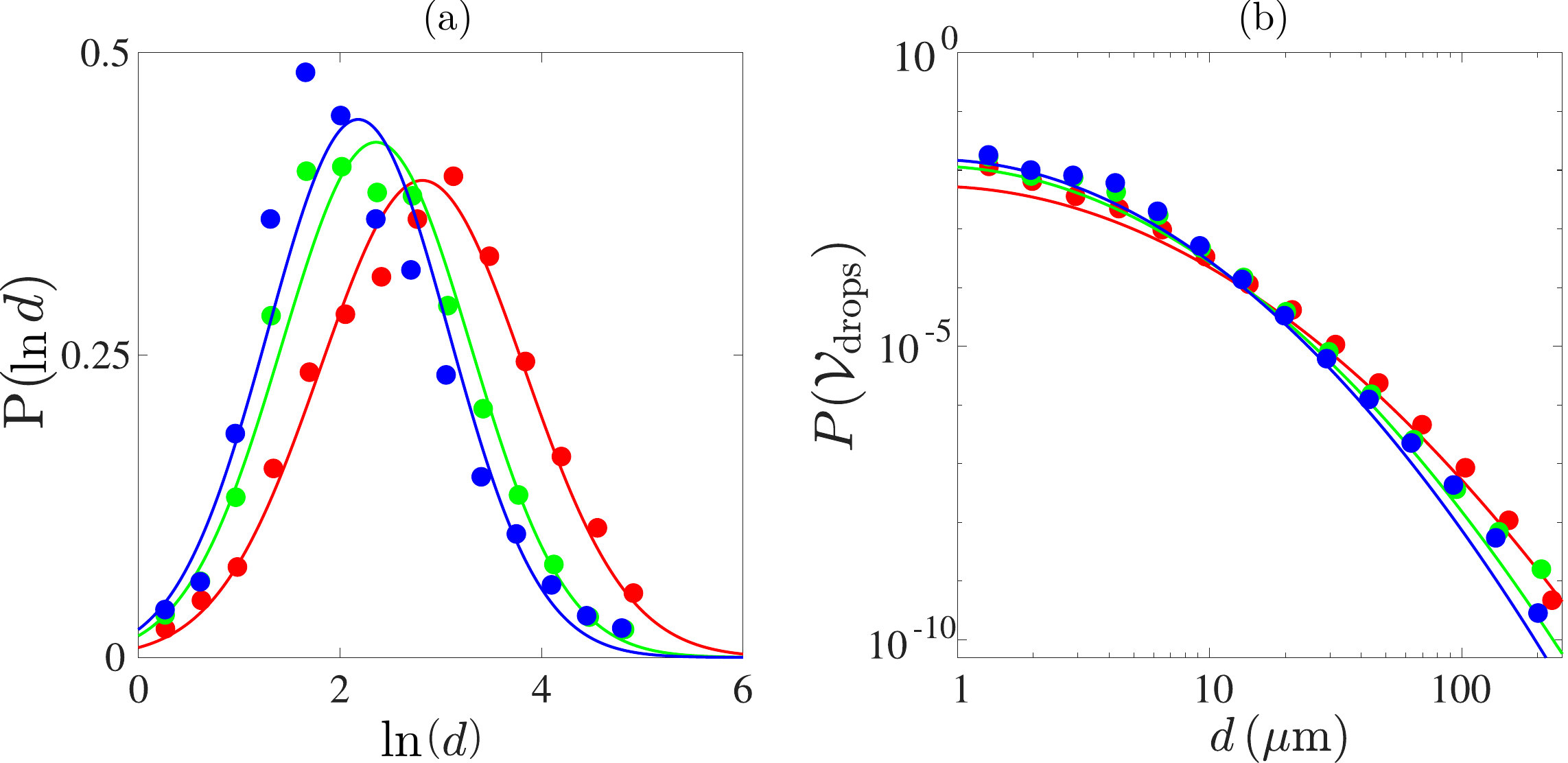}\\
\caption{Droplet size distribution measured at 10 cm away from the the exit of the channel. (a) Size statistics measured for different film viscosity \textcolor{red}{$\bullet$} = 5\,cst, \textcolor{green}{$\bullet$} = 25\,cst and \textcolor{blue}{$\bullet$} = 105\,cst. For these measurements the flow velocity and film thickness was fixed at $U = 30\,\mathrm{m/s}$ and $\mathcal{H}_\mathrm{f} = 1\,\mathrm{mm}$. Solid lines are a \emph{log-normal} fit to the data. (b) Volume based size statistics measured for different film viscosities. Solid lines are corresponding \emph{log-normal} fits to the data. }
\label{fig:fig7}
\end{figure}

We find that the resultant droplet size statistics, measured at the exit of the \emph{cough-machine}, are very-well captured by a \emph{log-normal} distribution, see Fig.\,\ref{fig:fig7}.
This behaviour is confirmed for different film viscosities and shearing strengths.
Note that our measurements include droplets as small as 1\,$\mu$m that originate from the unsteady rim retraction dynamics.
Crucially, our measurements of the volume size distribition (Fig.\,7b) are inline with previously reported droplet statistics of bioaerosols exhaled during expiratory activities \cite{han2013characterizations, morawska2009size, johnson2011modality}.
Previous investigations report the presence of both unimodal and bimodal \emph{log-normal} size distributions in bioaerosols exhaled by humans.
Crucially, this variation is found to be dependent on the physiological characteristics of the test subjects.
Therefore, a detailed investigation is warranted to quantify the influence of physiological characteristics on the physical (rheological) properties of the mucus-layer which in turn influences the droplet size statistics in exhaled bioaerosols.
Further, we must also point that our results are in contrast with other droplet size distributions suggested for bag-mediated fluid fragmentation processes \cite{villermaux2009single}. 
Notably, it has been proposed that in bag-mediated atomization scenarios, the aggregation and coalescence process of corrugations along the liquid filaments bordering inflated structures primarily selects the droplet size distribution.
Whereas, here we argue that a significant volume of atomized liquid is associated with small drop fragments that originate from the bursting of inflated liquid sheets.
In addition, we must also point out that due to the fact that the droplet size statistics are measured at a distance of 10 cm from the exit of the channel, the exact population of larger droplets formed due to the breakup of bordering filaments may not be well represented in our data, since large droplets are known to ballistically fall out of the expelled turbulent puff \cite{wells1936air, chong2021extended}, thus providing a bias in the measured droplet size statistics.

\begin{figure}
\centering
\includegraphics[clip, trim=0cm 0cm 0cm 0cm, width=0.8\textwidth]{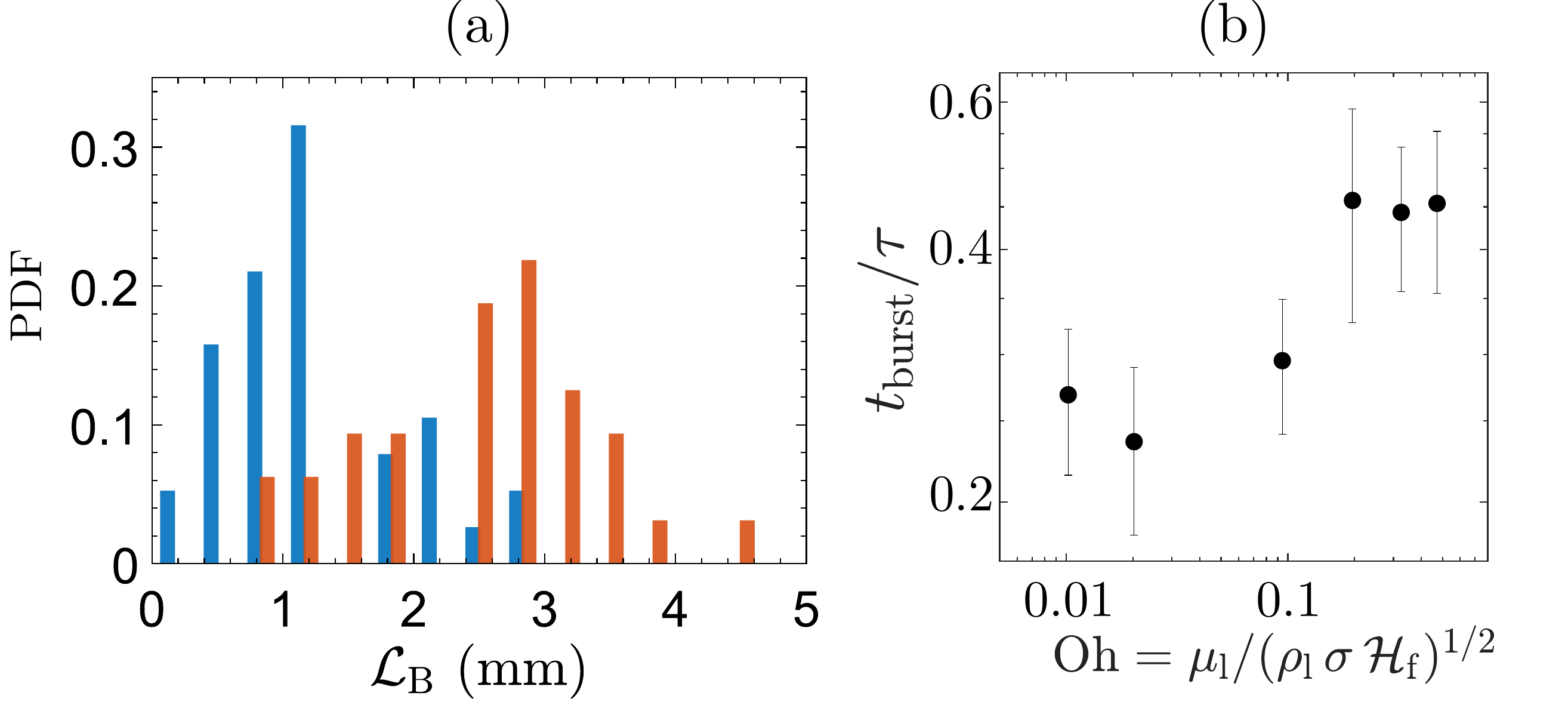}\\
\caption{(a) Histograms comparing the length of inflated morphologies  measured at the time of bursting for different film viscosity (blue) 5 cst and (red) 75 cst. Stabilizing effect of viscosity prolongs the lifetime of inflated structures. Accordingly, we measure delay in burst time of liquid bags with increase in film viscosity (b).}
\label{fig:fig8}
\end{figure}

\begin{figure}
\centering
\includegraphics[clip, trim=0cm 0cm 0cm 0cm, width=0.45\textwidth]{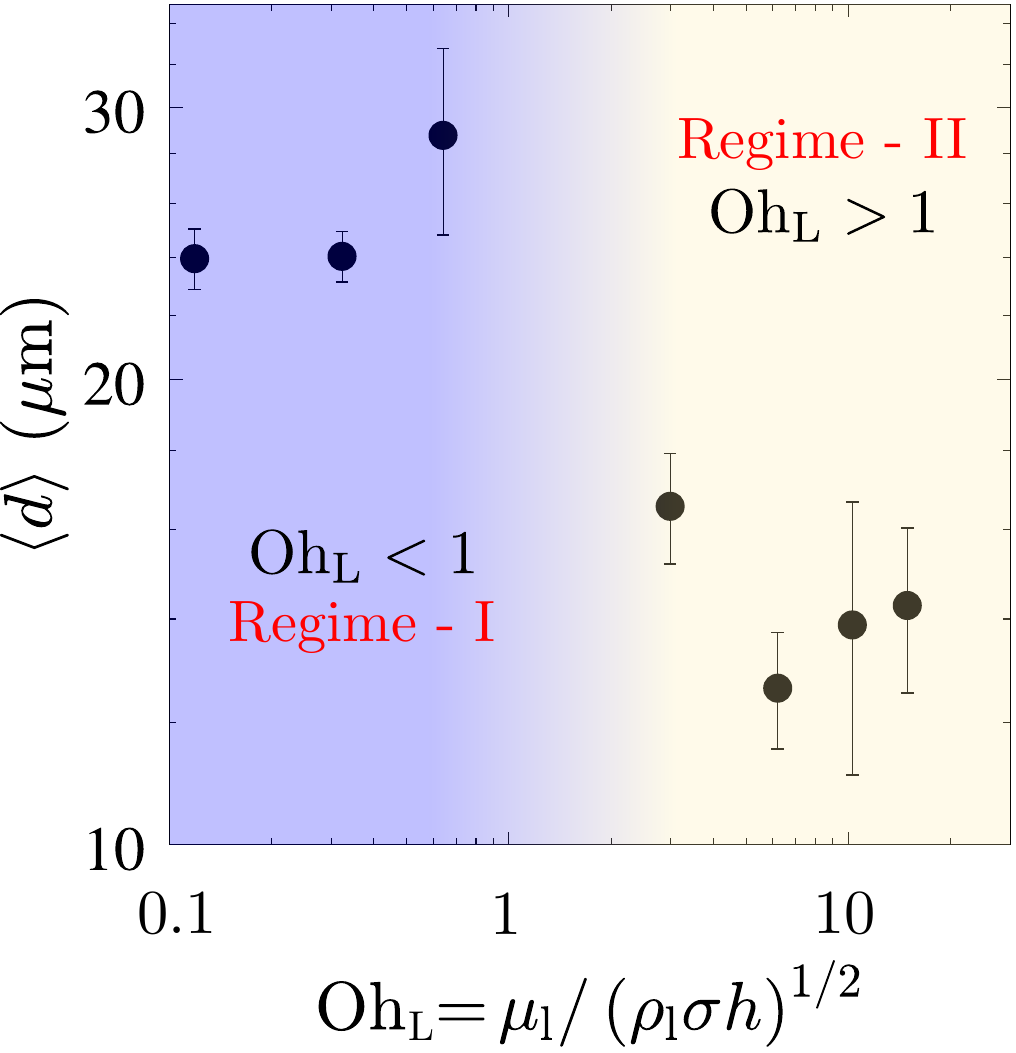}\\
\caption{Experimentally measured variation in the mean droplet size $\left< d \right>$ with local Ohnesorge number $Oh_\mathrm{l} = \mu/(\rho\,\sigma\,h)^{1/2}$. The length scale in $Oh_\mathrm{l}$ is the typical bag-thickness $h \sim 1\,\mu$m. In Regime-\Romannum{1}, $Oh_\mathrm{l} < 1$ a balance between inertia and capillary forces decides the eventual fate of a bag, causing it rupture early, whereas in Regime-\Romannum{2}, $Oh_\mathrm{l} > 1$ viscous effects enhance the stability of the thin sheets, causing it to burst at later times, thus producing smaller droplets.}
\label{fig:fig9}
\end{figure}

Our experiments also reveal an important role of fluid viscosity in the breakup of liquid bags.
We find that an increase in the film viscosity enhances the stability of inflated structures, resulting in the formation of deeper and wider liquid bags; Figs.\,\ref{fig:fig8}a and 4d.
Crucially, the enhanced stability of these inflated structures also means that the bursting time ($t_\mathrm{burst}$) of a bag is delayed.
In Fig.\,\ref{fig:fig8}b, we show the transition towards longer bursting times measured in experiments with an increase in film viscosity.
Notably, this result is inline with a delay in characteristic breakup times measured for bag-mediated breakup of secondary atomization of viscous droplets \cite{hsiang1992near}.
Crucially, we find that the enhanced stability of the bag directly affects the mean size of droplet fragments produced as a result of the atomization process.
As shown in Fig.\,\ref{fig:fig9}, for higher film viscosity the mean size of droplet fragments reduces significantly, almost by a factor of two.
Interestingly, this transition in mean droplet size occurs at the threshold given by the local Ohnesorge number $Oh_\mathrm{L} = \mu_\mathrm{l}/(\rho_\mathrm{l}\,\sigma\,h)^{1/2} \sim 1$, indicating the existence of two distinct regimes where different physical processes control the eventual bursting of the inflated structure.
We suspect that in the regime $Oh_\mathrm{L} < 1$, it is a balance between inertia and capillarity that determines the lifetime of an inflated structure, whereas for $Oh_\mathrm{L} > 1$, viscous effects stabilize the accelerating bag.
We also note that a similar stabilizing effect of fluid viscosity on the stability of thin films, in the context of surface bubbles, has been reported recently in Ref. \cite{lorenceau2020lifetime}.
However, a detailed investigation is further needed to develop a comprehensive understanding of the stability of inflated structures in these two regimes discussed above, and is out of scope of the current work.

In summary, motivated by the generation of bioaerosols during violent breathing manoeuvres, we have analyzed physical processes associated with the atomization of a thin film subject to shearing airflow in a closed geometry.
Our experiments and numerical simulations illustrate that the overall fragmentation process of a thin film is mediated by the formation of hollow bag-like structures. 
In contrast to previous investigations, we show that the cascade from an inflated bag to small droplets is controlled by the unstable motion of retracting liquid rims that border the holes puncturing the bag.
Crucially, we find that small droplets generated from such retracting liquid rims collectively carry a large liquid volume than previously thought. 
Accordingly, an effective infection control strategy must emphasize on the reduced transport of such small droplets. 
Importantly, our measurements of the resultant droplet size distribution match well with the some of the previously reported droplet statistics for a real cough.
However, the influence of a key feature of bioaerosol generation, i.e., viscoelasticity of the muco-salivary fluid, on droplet statistics is yet to be explored. 
We also reveal the stabilizing effect of fluid viscosity in the overall fragmentation process that eventually leads to the generation of smaller droplets. 
Finally, we envisage our investigation to motivate future research addressing the generation of bioaerosols, especially in the context of complex biological fluids.

\section*{acknowledgments}
The authors thank Gert-Wim Bruggert for the technical support in building the experimental setup, and Pim Wassdorp for the help in PDA measurements. We acknowledge the funding by Max Planck Center Twente, NWO and from the ERC Adv. Grant DDD 740479.

\bibliographystyle{prsty_withtitle}
\bibliography{bibliography}

\end{document}